# Photonic and plasmonic guiding modes in graphene-silicon photonic crystals

*Tingyi Gu[1,‡], Andrei Andryieuski[2,‡], Yufeng Hao[3], Yilei Li[4], James Hone[3], Chee Wei Wong[3], Andrei Lavrinenko[2], Tony Low[1*] and Tony F. Heinz[1,4]*

[1] Department of Electrical Engineering, Columbia University, New York, NY 10027

[2] DTU Fotonik, DTU, Oersteds pl. 343, Kongens Lyngby, Denmark 2800

[3] Department of Mechanical Engineering, Columbia University, New York, NY 10027

[4] Department of Physics, Columbia University, New York, NY 10027

**ABSTRACT.** We report systematic studies of plasmonic and photonic guiding modes in large-area chemical-vapor-deposition-grown graphene on nanostructured silicon substrates. Light interaction in graphene with substrate photonic crystals can be classified into four distinct regimes depending on the photonic crystal lattice constant and the various modal wavelengths (i.e. plasmonic, photonic and free-space). By optimizing the design of the substrate, these resonant modes can magnify the graphene absorption in infrared wavelength, for efficient modulators, filters, sensors and photodetectors on silicon photonic platforms.

TEXT

*Introduction.* Continual miniaturization of silicon photonics components has enabled future applications in densely integrated communication and computing systems[1]. The high refractive index and low absorption in silicon allow for low-loss waveguiding from terahertz to telecommunication bandwidth, but the bandgap of 1.1 eV renders silicon inefficient as photo detection material in the long wavelength range. Hence, photonic systems based on alternative materials such as III-V semiconductors[2] and graphene[3] are being sought for in those applications[4]. In particular, graphene is promising since it accommodates mid-infrared plasmons[5,6] with strong light confinement down to 1/100 of the free-space wavelength $\lambda_0$, in conjunction to its unique tunability of electrical conductivity, and thus allowing for active devices[7] not possible with conventional metal plasmonics. Graphene plasmons has already been experimentally observed[7–12] and explored for terahertz and infrared absorption and modulation, photodetection and chemical sensing[13–18]. Furthermore, graphene can be easily integrated with silicon photonics components for more efficient light management schemes[14,19].

One of the challenges for surface plasmon polaritons excitation in graphene is the phase (optical $k$-vector) mismatch with the incident electromagnetic wave. Graphene nanostructures, such as one-dimensional (1D) nanoribbons array[11,12,20] or two-dimensional (2D) rectangular resonator array[21], dots[7] and antidots lattices[22,23] and plasmonic crystals[24] are straightforward means for light coupling to graphene plasmons. However, plasmons excited in this way are limited by severe damping pathway due to the patterned edges with atomic scale roughness[11,25]. It would, therefore, be more optimal to form light-plasmon coupler gratings by patterning the substrate instead and leaving the graphene layer pristine in order to preserve the excellent electronic properties of graphene.

1D grating etched onto a silicon substrate with a graphene overlaid on top has been studied theoretically for mid-IR plasmons excitation[26] and subsequently investigated experimentally in a hexagonal 2D grating configuration[27]. These configurations exploit the difference in the effective mode index between plasmons on silicon-graphene-air and suspended air-graphene-air interfaces, in addition to the dynamic conductivity contrast of supported graphene to suspended region[28]. At the same time, photonic modes within the silicon grating can also be excited. These photonic modes can also be used for an increased absorbance in graphene, and has been studied recently[29–33]. In this work, we explore various regimes of light-matter interactions in the infrared frequencies in graphene integrated with a silicon photonic crystal membrane. We demonstrate experimentally light coupling to the graphene plasmonic modes and the membrane photonic guiding modes, and theoretically explores their physical mechanisms and design space responsible for the crossover between these two regimes.

*Classification.* Figure 1a depicts a scanning electron micrograph of graphene on a photonic crystal membrane. The different regimes of light coupling schemes can be categorized based on the structure feature size (such as the lattice constant, *a*) and effective wavelengths for plasmonic ($\lambda_{plasm}$) and photonic ($\lambda_{photon}$) modes and incident (in the considered for from the free space) wavelength ($\lambda_0$) (typically, $\lambda_{plasm} < \lambda_{photon} < \lambda_0$) as follows:

1. $a < \lambda_{plasm}$: **Metamaterial regime**. In order to excite a guided wave efficiently under the normal incidence, the distance between scatters in a periodic array should be equal to an integer number of effective wavelengths of the guided mode. Therefore, we may consider the system with $a < \lambda_{plasm}$ as a graphene plasmonic metamaterial (or metasurface), and no guided surface waves can be excited due to the pronounced phase mismatch. The incident wave can only be reflected, transmitted and absorbed in graphene (absorption in silicon is low in the considered spectral range). Structured silicon substrate can be considered as a layer of anisotropic dielectric[34].

2. $\lambda_{plasm} < a < \lambda_{photon}$: **Plasmonic regime**. Together with transmission and reflection we may observe coupling of the incident wave to plasmonic modes at certain frequencies (coupling will occur to the waves with the propagation constant $\beta = qk_0 = \frac{2\pi}{a}m$, where $q$ is the effective mode index, $k_0 = 2\pi/\lambda_0$ is the wavenumber and $m$ is an integer number). The incident wave couples to graphene plasmons which are tightly confined along the graphene layer. Partially their energy is absorbed alongside and partially out-coupled back to the free space.

3. $\lambda_{photon} < a < \lambda_0$: **Photonic regime**. In addition to the aforementioned channels the incident wave can couple to the photonic modes of the membrane, and its power is absorbed by graphene, which acts mostly as an absorbing layer. Photonic modes number and properties depend not only on the grating lattice constant, but also on the thickness of the silicon layer. Since we assume light coupling from a normally incident plane wave, we deal with the photonic modes that are normally considered as leaky.

4. $\lambda_0 < a$: **Diffraction grating regime**. The periodic structure becomes a diffraction grating and the incident power will be distributed to the higher diffraction orders in addition to the photonic and plasmonic modes excitation.

It is worth mentioning that for a sufficiently large lattice constant $a$ all these regimes are possible simultaneously, though, usually in different frequency ranges. The metamaterial regime for absorbers is well-known[35], while the diffraction grating regime is hardly useful for absorbers, since the incident power is predominantly scattered into the free space. In this work we will explore experimentally and theoretically the most interesting plasmonic and photonic modes excitation regimes (see artistic illustration in Figure 1a).

*Experiment.* The 2D hexagonal silicon photonic crystal membranes were fabricated on a 250 nm thick silicon-on-insulator device layer via optimized 248 nm deep ultraviolet lithography and etching for reduced disorder scattering[36]. The lattice constant of the photonic crystal is 415 nm with the hole radius varying from 80 nm to 140 nm in step of 10 nm, corresponding to the silicon filling fraction of 0.87-0.59, respectively. The sacrificial release of the supporting silicon oxide buffer layer was performed by wet etching, resulting in 0.5 μm air gap between photonic crystal membrane and silica (see Figure S1 Large-area graphene films were grown on Cu substrates by chemical vapor deposition[37]. The film thickness and crystalline quality were characterized with Raman spectroscopy (data not shown here) first and then transferred onto the patterned substrate via a PMMA-assisted methods (Figure 1a-b)[38]. Atomic force microscopy measures the 0.57 nm height difference between the supported and suspended graphene, with different surface roughness (Figure 1c). Good contrast in the surface stiffness of the suspended and supported parts of graphene is used to map the periodic structure (Figure 1d). The formation of graphene-silicon anti-dots arrays with fixed lattice constant $a$ and increasing hole radius $r$ is identified by the surface stiffness mapping [Figure 1d (i-vi)] and diffraction pattern via two dimensional Fourier Transformation [Figure 1d (i'-vi')].

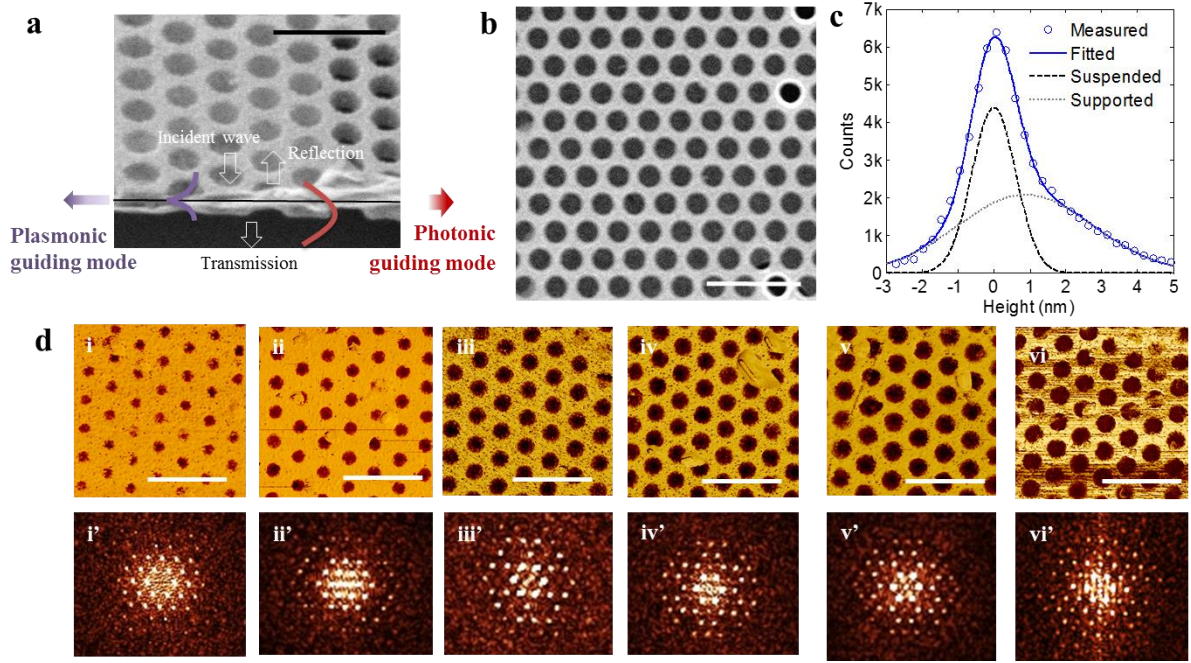

**Figure 1. Suspended graphene superlattice:** (a) SEM image with an artistic view of the plasmonic and photonic modes excited by the normally incident plane wave, (b) top view of SEM devices, (c) the histogram of the AFM measured topology. The experimental peak is decomposed in the suspended part (FWHM of 1.72 nm) and supported part (FWHM of 5.31 nm), with 0.57 nm difference. (d) AFM measured surface stiffness of graphene arranged on a silicon photonic crystal membrane with different disk radius. (i) $r$ = 90nm, (ii) 100nm, (iii) 110nm (iv) 120nm (v) 130nm (vi) 140nm and corresponding 2D spatial Fourier transformation (i'-vi', respectively). The lattice constant $a$ = 415 nm, the scale bars equal to 1 μm.

In order to identify graphene's contribution to the optical spectrum modification on the patterned substrate, we selectively etched away a half of graphene by the oxygen plasma (see Supporting information, Figure S5a). The graphene coverage on the substrate is characterized by Raman spectroscopy (Figure S5b), as a free-standing graphene monolayer has a stronger Raman signal compared to the supported part[39].

For broadband FTIR characterization, the sample was mounted on VERTEX FTIR spectrometer (Bruker optics). Broadband light from a thermal source (near to far-IR range) normally incident onto the sample through a spatially selective window of 100 μm by 20 μm (outlined as the white dashed rectangle in Figure S1a). Due to large material losses in silica around 1000 cm$^{-1}$, transmission measurements are inaccurate, so we measured reflectance $R$ from the graphene-covered and $R_0$ from non-covered reference and calculated extinction ratio $= 1 - R/R_0$. The extinction ratio, therefore, quantifies the contribution due to the presence of the graphene layer.

*Simulation.* Reflectance from the graphene-covered and reference structures (Figure S1) were simulated in CST Microwave Studio[40] with the time-domain solver assuming periodic boundary conditions. Material properties of silicon and silica (see Supporting information, Figure S2a-b) were taken from the Refs.[41,42], respectively. Graphene was modeled as a $\Delta = 1$ nm thick layer with the effective dielectric permittivity $\varepsilon_G(\omega) = 1 + i\sigma_S(\omega)/\varepsilon_0\omega\Delta$, where $\omega$ is the angular frequency and $\sigma_S(\omega)$ is the conductivity of graphene[43]

$$\sigma_S(\omega) = \frac{2e^2 k_B T}{\pi \hbar^2} \ln\left[2\cosh\left(\frac{E_F}{2k_B T}\right)\right] \frac{i}{\omega + i\gamma} + \frac{e^2}{4\hbar}\left[H\left(\frac{\omega}{2}\right) + \frac{4i\omega}{\pi}\int_0^\infty dx \frac{H(x) - H\left(\frac{\omega}{2}\right)}{\omega^2 - 4x^2}\right],$$

where $H(x) = \frac{\sinh\frac{\hbar x}{k_B T}}{\cosh\frac{E_F}{k_B T} + \cosh\frac{\hbar x}{k_B T}} = \frac{1}{2}\left[\tanh\left(\frac{\hbar x + E_F}{2k_B T}\right) + \tanh\left(\frac{\hbar x - E_F}{2k_B T}\right)\right]$.

We used the electrochemical potential $E_F = 0.3$ eV (0.25-0.35 eV is a typical value of electrochemical potential experimentally measured with terahertz spectroscopy for transferred non-intentionally doped graphene[44,45]) and damping rate $\gamma = 2.0 \times 10^{13}$ s$^{-1}$ (a realistic value close to experimentally measured $\gamma = 2.3 \times 10^{13}$ s$^{-1}$ for large area CVD graphene[46]) for numerical simulations (see Supporting information, Figure S2c).

*Plasmonic regime.* In the lower frequency mid-IR region, we expect light-matter interactions dominated by plasmonic modes of graphene. The silicon membrane is much thinner as well as the lattice constant *a* is much smaller than the wavelength so the membrane can be considered as an effectively homogenous dielectric layer for the optical (but not plasmonic) waves (see its effective index in Figure S2d). As a characteristic value, the effective wavelengths for the plasmons at the frequency 1000 cm$^{-1}$ (see Supporting information Figure S3) are 352 nm and 54 nm and the propagation lengths - 264 nm and 41 nm for the suspended and supported graphene, respectively. The size of the resonator should be nearly equal to an integer number of effective wavelength. For the hole radius $r = 100$ nm, the size of the silicon area between the edges of nearest neighboring holes is 215 nm. Based on these simple considerations, we expect to excite the second order plasmonic mode (with the effective wavelength about 112 nm around 700 cm$^{-1}$) at silicon-graphene-air interface (the first order mode lies below 500 cm$^{-1}$). The air-graphene-air plasmons can be excited at larger frequencies (first order resonance at around 1200 cm$^{-1}$) though still smaller than the frequencies of interband transitions. Due to small propagation length of plasmon it is expected that the quality factor of the resonator is low and the resonance broad.

This expectation is confirmed by the experiment and simulation as shown in Figure 2 a-b. The plasmonic peak in the range below 1000 cm$^{-1}$ experiences a gradual blue-shift with increasing hole's radius (decreasing the size of graphene supported on silicon). This basic trend is consistent with the fact that the silicon-graphene-air plasmonic modes are excited. The inset in Figure 2b shows the quickly spatially oscillating in-plane electric field distribution at the associated resonance, showing the second order standing plasmonic mode in the supported graphene. The characteristic peak around 1100 cm$^{-1}$ is well known, and is due to maximum material absorption in silica due to the associated infrared phonon mode. The presence of strong absorbance peak in silica prohibits observation of air-graphene-air plasmonic resonances otherwise expected in the range above 1200 cm$^{-1}$.

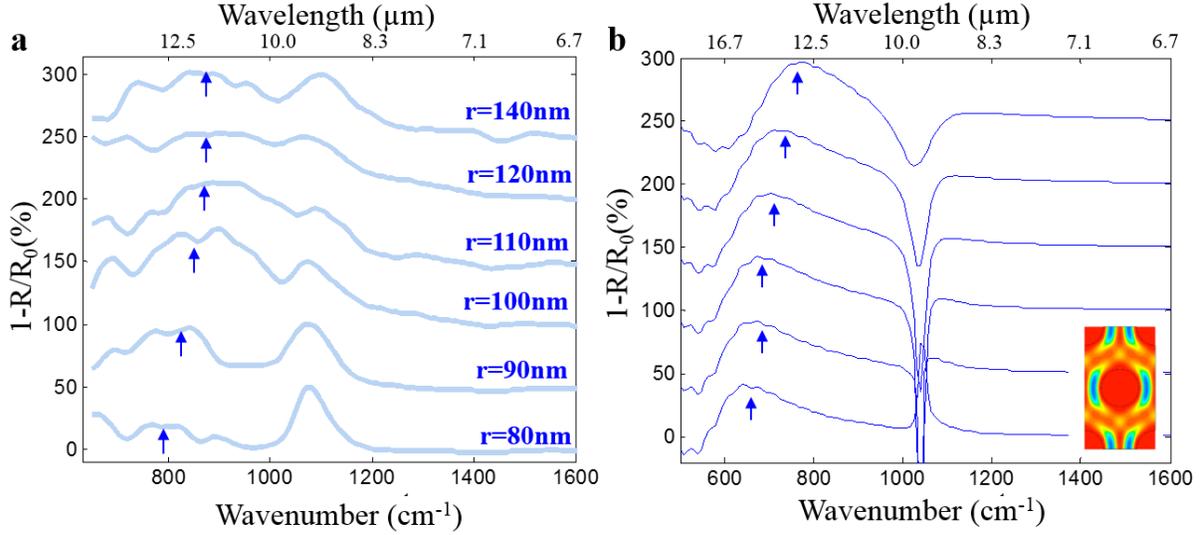

**Figure 2.** Extinction spectrum for graphene on a photonic crystal in the plasmonic regime in mid-IR range. (a) Measured and (b) simulated extinction spectra of graphene covering 250 nm thick photonic crystal membranes with fixed lattice constant and increasing hole radius ($r$ = 80, 90, 100, 110, 120 and 140 nm). Vertical cumulative offset of 50% is added for clarity. The peak around 1100 cm$^{-1}$ comes from the silica material absorption. Inset: top view of the in-plane electric field profile of the mode corresponding to the peak.

*Photonic regime.* For the considered electrochemical potential $E_F$ = 0.3 eV the interband transitions occur for the frequencies above 4830 cm$^{-1}$, thus in the higher-frequency range 5500 cm$^{-1}$ – 7700 cm$^{-1}$ graphene is simply an absorptive material with constant $\sigma_S(\omega) = \sigma_0 = e^2/4\hbar$. Figure 3a shows the measured extinction spectrum in reflection, $ER = 1 - R/R0$, showing a distinctive resonance corresponding to the first Fabry-Perot photonic resonance (compare the frequency position of the resonance to the kink in effective index of the membrane in Figure S2d) with maximal absorbance systematically blue shifted with the increasing hole radius. These spectra were found to be well-fitted by Fano line shapes.

Figure 3b shows the corresponding simulation of our device, which consists of several relatively thick layers (250 nm thick silicon photonic crystal membrane, 500 nm thick air gap, 1500 nm thick silica and then thick silicon substrate, see Figure S1) thus forming a multi-section cavity with multiple resonances in the range of the FTIR spectrometer (see the Supplementary information). However, there is only one resonance in the considered frequency range 5500-7700 cm$^{-1}$ that is responsible for the observed extinction ratio, and reproduced the blue shifts with increasing holes' radius following the same trend as in the experimental spectra. The asymmetric Fano lineshape is less apparent in simulations due to the less pronounced resonance features in the extinction spectrum related to interference with another resonance at larger frequencies. . The electric field distribution is illustrated in the inset of Figure3b, which shows that the field concentrates mainly around graphene, hence maximizing its interaction with the photonic mode.

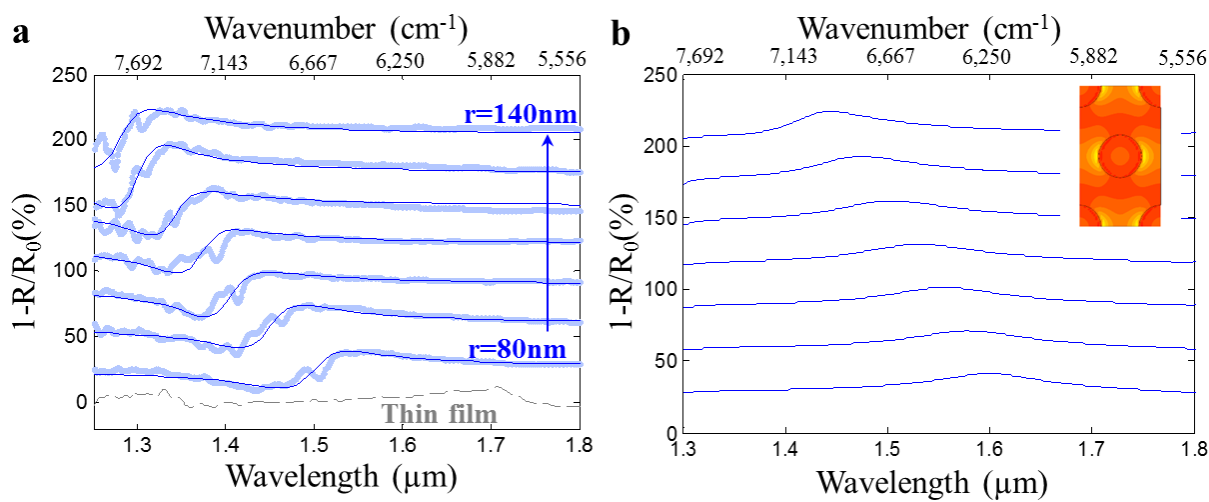

**Figure 3.** Graphene extinction spectrum exhibiting the coupling with photonic resonance in the near-IR range. (a) Measured extinction spectra of graphene covering 250 nm-thick silicon film

(grey dashed curve) and photonic crystal membranes with fixed lattice constant $a$ = 415 nm and increasing hole radius from 80 to 140 nm (light blue). Solid blue curves are Fano resonance curve fitting. Vertical cumulative offset of 30% is added for clarity. (b) Numerically simulated extinction for the corresponding measured photonic crystal membranes. Inset: vertical electric field amplitude profile for the structure at resonant frequency, dashed line shows graphene positioned in the field maximum.

*Maximizing absorbance.* There are many ways to maximize absorbance of an optical structure[47]. In subsequent discussion, we consider graphene on a simplified one-dimensional photonic crystal (Figure 4) with a period $a$, thickness $t$ and air hole width $w$. Absorbance dependence on the structural parameters is discussed in the Supporting information S1.3. The electric fields at resonant frequencies represent the whole "zoo" of possible excitations (Figure 4a, grey curve), including silicon-graphene-air and air-graphene-air modes, mixed modes and photonic modes see the inset for the field mapping.

In case coupling to the photonic modes, a reasonable approach is to optimize the geometrical parameters so that the photonic modes can propagate sufficiently long distance before out-coupling. In this way the performance of graphene absorber will be improved. For example, increasing the period $a$ to 500 nm for the thick silicon membrane of $t$ = 500 nm leads to absorbance increase from a few percent to 18% (See Supplementary Information, S1.3).

Even though the grating allows for in-coupling of incident wave to photonic modes, being a reciprocal device, it also allows for out-coupling to the free space. A good solution for avoiding the mentioned out-coupling is to couple first the light from the free-space to a photonic waveguide with the help of a grating (coupling efficiency can be very high, up to -0.5 dB)[48] and then to absorb the radiation with graphene upon wave's propagation along the waveguide. Such solution does not give any advantage for plasmonic regime (at least for large damping rate $\gamma$), since plasmonic modes experience fast decay without being out-coupled to the free space.

Another way to increase the absorbance both in plasmonic and photonic regimes is to use a back reflector (mirror) – see the red curve in Figure 4a. This technique is known for several decades as the Salisbury screen. he real part of graphene's conductivity should be appropriate that can be reached by increasing Fermi energy or employing a few graphene layers- An appropriate thickness of the dielectric spacer allows for phase matching, thus placing graphene in the maximum of the standing wave[35]. For example, adding a metallic mirror with an air spacer $t_{air}$ = 1600 nm below the $t$ = 100 nm thick membrane with graphene ($E_F$ = 0.7 eV) allows for absorbance increase from 18% to 87% as shown in Figure 4a. Absorbance for each plasmonic mode experiences maximum for its specific air gap thickness (Figure 4b).

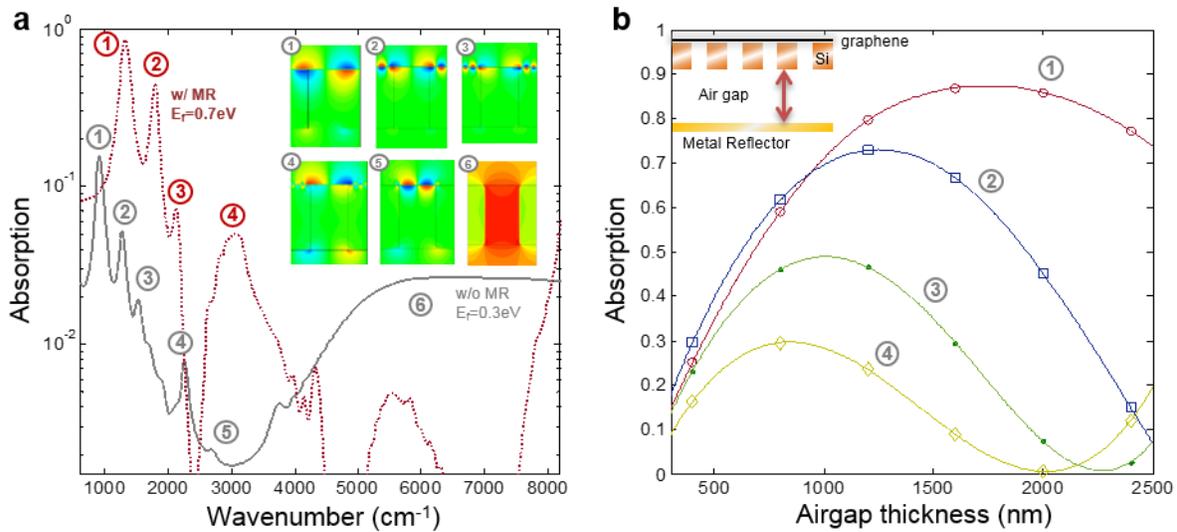

**Figure 4**. Optimization of absorption enhancement by plasmonics in mid-IR range. (a) Absorbance of graphene ($E_F$ = 0.7 eV) on silicon photonic crystal membrane ($a$ = 100 nm, $w/a$ = 0.1, $t$ = 100 nm) with (red dotted curve) and without (grey solid curve) back mirror reflector (air gap is 1600

nm). The field profiles at certain frequencies are shown in the inset. Maximum absorption of 87% is achieved by increasing Fermi level in graphene to 0.7 eV and setting the air gap thickness to 1600nm (see the inset of b) (b) Absorbance peak intensity of four plasmonic modes (1-4, see panel a) versus the air gap thickness.

*Discussion.* In summary, graphene combined with a photonic crystal membrane is an extremely interesting system, which presents a rich playground of light-matter interactions. Most importantly, the SOI based membrane is fully compatible with the silicon based CMOS fabrication technology. We present different regime for IR light absorption enhancement in a graphene-over-silicon membrane structure, where the membrane is patterned with a two-dimensional photonic crystal lattice. The enhancement is a direct consequence of light coupling to plasmonic and photonic modes. The absorption spectra are studied through systematic variation of the underlying photonic crystal hole radius, and the observed results are consistent with theoretical simulations.

We classified various regimes of light interactions with the structured composite, namely, metamaterial, plasmonic, photonic and diffraction grating, according to their effective wavelengths $\lambda_{plasm} < \lambda_{phot} < \lambda_0$. Each regime can be suitable for different applications. The metamaterial regime, mainly in the terahertz-mid-IR range can be used for tunable absorbers, filters and modulators. The plasmonic regime in mid-IR can be used, in addition to the aforementioned applications, for sensing due to the extreme confinement of graphene plasmons close to the surface. The photonic regime can be also used for photodetection, since absorption of photons in graphene may generate a photocurrent.

## ASSOCIATED CONTENT

**Supporting Information**. Simulated materials and structures; graphene plasmons properties; absorbance in graphene-covered 1D silicon subwavelength grating; Raman characterization of graphene-on-photonic crystal structure; effective index calculation and reflectivity of multilayer structure; Fano fit to the substrate guiding modes in near IR region. This material is available free of charge via the Internet at http://pubs.acs.org.

## AUTHOR INFORMATION

**Corresponding Author**

*E-mail: tonyaslow@gmail.com, tlow@umn.edu

**Notes**

The authors declare no competing financial interest.

**Author Contributions**

T.G. and A.A. contributed equally.

## ACKNOWLEDGMENT

A. A. acknowledges financial support from the Danish Council for Technical and Production Sciences through the GraTer project (Contract No. 0602-02135B). The authors acknowledge experimental support of AFM from C. Forsythe in Prof. P. Kim's group. T. Gu thanks discussions with Z. Huang and C. Li at Hewlett-Packard Laboratories.

TOC GRAPHIC

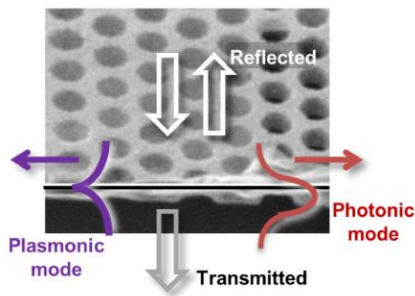

# Supporting information for

# Photonic and plasmonic guiding modes in graphene-silicon photonic crystals

*Tingyi Gu1†, Andrei Andryieuski2†, Yufeng Hao3, Yilei Li3, James C. Hone4, Chee Wei Wong4, Andrei Lavrinenko2, Tony Low1\* and Tony F. Heinz1,3*

[1] Department of Electrical Engineering, Columbia University, New York, NY 10027

[2] DTU Fotonik, DTU, Oersteds pl. 343, Kongens Lyngby, Denmark 2800

[3] Physics Department, Columbia University, New York, NY 10027

[4] Department of Mechanical Engineering, Columbia University, New York, NY 10027

**S1. Theoretical**

**S1.1. Simulated materials and structures**

The simulated structure (Figure S1) corresponds to the fabricated one. Refractive indices of silicon[1] and silica[2] are shown in Figure S2a-b, respectively. Graphene's conductivity in the random phase approximation can be calculated [3] as

$$\sigma_S(\omega) = \frac{2e^2 k_B T}{\pi \hbar^2} \ln\left[2\cosh\left(\frac{E_F}{2k_B T}\right)\right] \frac{i}{\omega+i\gamma} + \frac{e^2}{4\hbar}\left[H\left(\frac{\omega}{2}\right) + \frac{4i\omega}{\pi}\int_0^\infty dx \frac{H(x)-H\left(\frac{\omega}{2}\right)}{\omega^2-4x^2}\right], \quad (S2)$$

where $e$ is elementary charge; $k_B$ is Boltzmann's constant; $T$ is absolute temperature; $\hbar$ Planck's constant; $\gamma$ damping rate and $H(x)$ is defined as:

$$H(x) = \frac{\sinh\frac{\hbar x}{k_B T}}{\cosh\frac{E_F}{k_B T}+\cosh\frac{\hbar x}{k_B T}} = \frac{1}{2}\left[\tanh\left(\frac{\hbar x+E_F}{2k_B T}\right)+\tanh\left(\frac{\hbar x-E_F}{2k_B T}\right)\right] \quad (S3)$$

Graphene's conductivity for $E_F = 0.3$ eV, $\gamma = 2 \times 10^{13}$ s$^{-1}$ and $T = 300$ K is shown in Figure 2c. If the interband contribution to the conductivity is negligible, that is the case for $\hbar\omega < 2E_F$ and the electrochemical potential is larger than thermal fluctuation energy $E_F > k_B T$, graphene's conductivity can be described with Drude formula

$$\sigma_S(\omega) = \frac{e^2 E_F}{\pi\hbar^2}\frac{i}{\omega+i\gamma} \quad (S4)$$

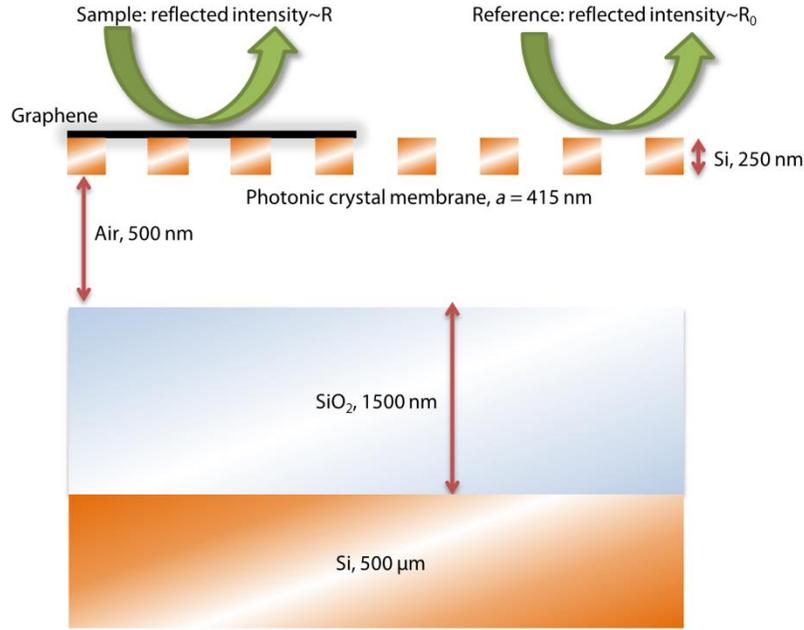

**Figure S1.** Simulated structure consists of a silicon photonic crystal membrane (thickness $d = 250$ nm, lattice constant $a = 415$ nm) suspended above silica (1.5 µm) on silicon substrate.

We restored the effective index of the suspended photonic crystal membrane from reflection and transmission simulation[4]. The effective index decrease with hole radii increase due a decreasing silicon filling fraction. A kink in the spectra shifting from 6000 to 7000 cm$^{-1}$ corresponds to the first Fabry-Perot resonance condition $d = \lambda_0/2n_{eff}$.

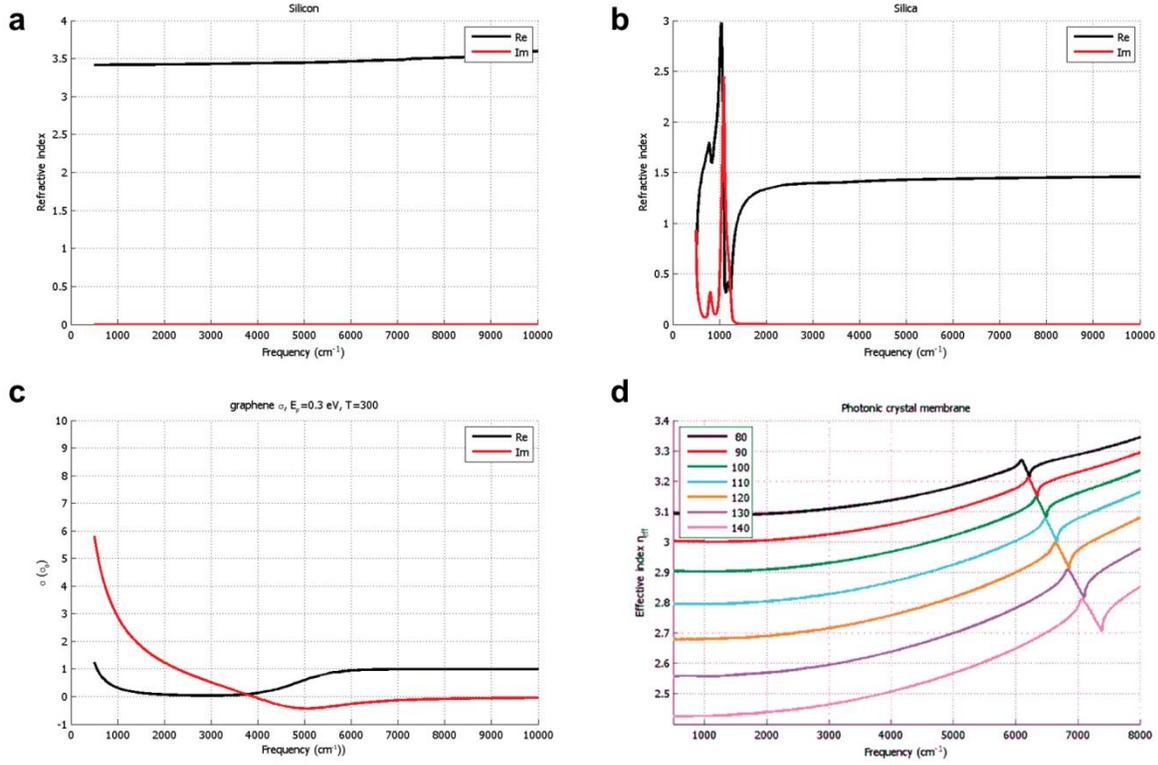

**Figure S2.** Effective index of the photonic crystal membrane depending on holes radii. The refractive index of silicon and silicon are given in ref. 1 and ref. 2 respectively. surface conductivity of graphene in units of $\sigma_0 = e^2/4\hbar$ ($E_F = 0.3$ eV, $\gamma = 2 \times 10^{13}$ s$^{-1}$, $T = 300$ K).

Simulations were done in CST Microwave Studio[5] with time-domain solver, rectangular mesh and effectively periodic (x-axis perfect electric conductor, y-axis perfect magnetic conductor, z- open) boundary conditions. Fine spatial discretization was needed for consistent results (for example, 1 nm thick graphene layer was discretized with the step of 0.1 nm). Typical time for one simulation was 6 hours on 12 CPUs (3 GHz), 48 GB RAM computer.

**S1.2 Graphene plasmons**

Dispersion of transverse magnetic (TM) plasmons in graphene layer placed between two dielectrics with the permittivities $\varepsilon_1$ and $\varepsilon_2$ is described by the dispersion equation[6]

$$\frac{\varepsilon_1}{\sqrt{q^2(\omega)-\varepsilon_1}} + \frac{\varepsilon_2}{\sqrt{q^2(\omega)-\varepsilon_2}} + i\sigma_S(\omega)Z_0 = 0 \tag{S1}$$

where $q = \beta/k_0$ is the normalized propagation constant or in other words effective mode index ($\beta$ is the propagation constant and $k_0 = \omega/c$ is the wavenumber in vacuum), $\sigma_S$ is the surface conductivity of graphene and $Z_0 = \sqrt{\mu_0/\varepsilon_0} = 1/c\varepsilon_0 = 120\pi [\Omega]$ is the free-space impedance.

Taking into account graphene's plasmons' large effective index $q \gg 1$ and typical dielectrics have permittivity in the range of 1-12, the dispersion relation can be simplified as:

$$q(\omega) = \frac{(\varepsilon_1+\varepsilon_2)\pi\hbar^2}{Z_0 e^2 E_F}(\omega + i\gamma) \tag{S5}$$

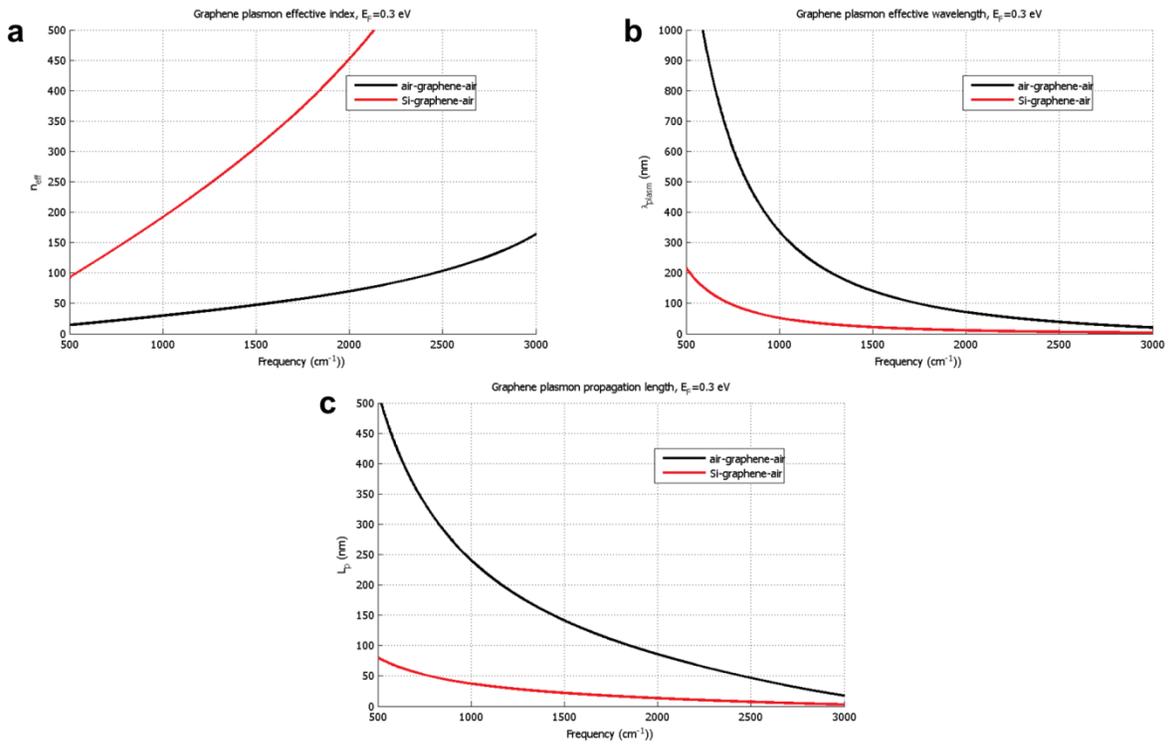

**Figure S3.** Effective index (a), effective wavelength (b) and propagation length (c) of plasmons in graphene suspended in air (black line) or supported on silicon (red line). Graphene's electrochemical potential is $E_F = 0.3$ eV and damping rate $\gamma = 2 \times 10^{13}$ s$^{-1}$.

Propagation constant (see the effective mode index for $E_F = 0.3$ eV and $\gamma = 2 \times 10^{13}$ s$^{-1}$ in Figure S3a) depends on the graphene's surrounding and thus the plasmons propagating on suspended graphene and graphene on silicon have different speed (effective index), thus forming plasmonic crystal or metamaterial (another way to form it is to structure graphene [7]). Typical plasmon wavelength (Figure S3b) as well as propagation length (Figure S3c) range from a few tens to a few hundreds nanometers in the frequency range of interest (500-3000 cm$^{-1}$).

**S1.3. Graphene covered 1D silicon subwavelength grating**

For a better understanding of the photonic and plasmonic regimes we simulated graphene on a simplified one-dimensional photonic crystal (see Figure S4) with a period $a$, thickness $t$ and air hole width $w$. The absorbance spectrum revealed a bunch of resonances. For a fixed period $a = 100$ nm and increasing air filling fraction $w/a$ (Figure S4a) we observe reduction of number of silicon-graphene-air resonances and their blueshift that is consistent with the previously observed excitation of the plasmons on supported part (with decreasing silicon size the resonator length decreases thus the resonant frequencies increase). Meanwhile the air-graphene-air plasmonic modes redshift (Figure S4a).

For the fixed w/a ratio and increasing grating period a (Figure S4b) there is observed a red-shift (resonator size becomes larger) with a trend resonance broadening. For the period a = 500 nm we can hardly distinguish more than one resonant peak whereas at a = 100 nm there are many. The reason for this is clear if we remember the typical propagation length for the supported graphene plasmons (see above) are less than 100 nm. For the resonator size of several hundreds of nm a plasmon excited at the edge of silicon decays before reaching the opposite side of the resonator, thus the resonance cannot be formed (in other words, the quality factor of the resonator is very low).

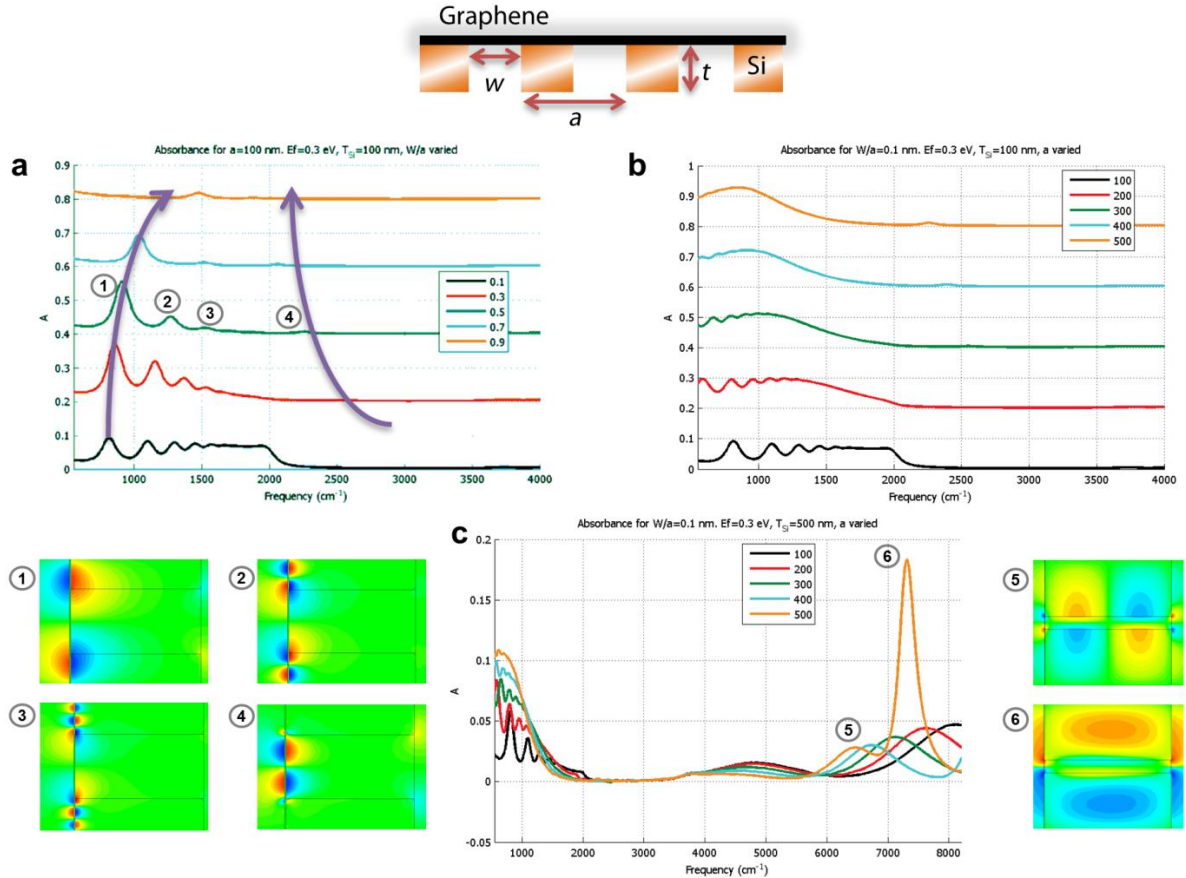

**Figure S4**. Simplified 1D photonic crystal consisting of silicon subwavelength grating and graphene ($E_F = 0.3$ eV) in air. (a) Absorbance for $a = 100$ nm, $t = 100$ nm and varying $w/a = 0.1$-$0.9$. Silicon-graphene-air plasmonic modes (insets 1-3) exhibits blue shift and reduction of number of modes with increasing $w/a$, whereas air-graphene-air mode redshifts (inset 4). (b) Absorbance for $w/a = 0.1$, $t = 100$ nm and varying period $a = 100 - 500$ nm. Silicon-graphene-air modes redshift and broaden with period $a$ increase. (c) Absorbance for a thick membrane $t = 500$ nm, $w/a = 0.1$ and varying period $a = 100 - 500$ nm. Plasmonic modes exhibit a similar behavior to the case (b) in the low frequnecies, whereas in the high frequencies photonic modes (insets 5 and 6) are observed.

In the previous cases (Figure S4a-b) the silicon membrane was assumed $t = 100$ nm thick. In low frequencies, in principle, there are no principal differences between thick and thin membranes (as soon as the membrane is still subwavelength). Small thickness does not allow for photonic modes excitation and guiding. If we increase the thickness to $t = 500$ nm plasmonic modes exhibit qualitatively the same behavior in the low frequencies, while in the high frequencies absorbance peaks corresponding to photonic modes appear (Figure S4c). The electric fields at resonant frequencies represent the whole "zoo" of possible excitations (Figure S4, insets 1-6), including silicon-graphene-air and air-graphene-air plasmonic and silicon grating photonic modes.

## S2. Experimental

### S2.1 Raman characterization of graphene-on-photonic crystal structure

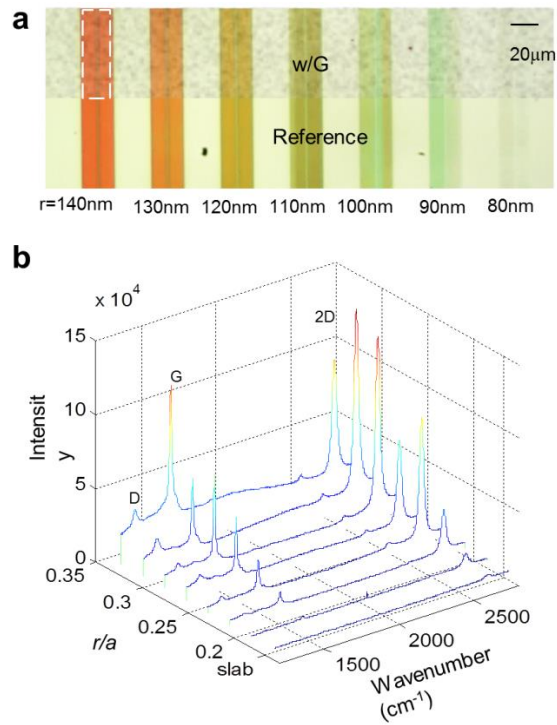

**Figure S5.** (a) Optical microscope image of device layout, with strips of silicon superlattice with hole radius $r$ decreasing from 140 to 80 nm in step of 10 nm (left to right). Graphene covers the shadowed area on top. The dashed square marks the region of FTIR window. (b) Raman spectrum for graphene superlattice with different radius versus lattice constant ratio $r/a$.

### S2.2 Effective index calculation and reflectivity of multilayer structure

The suspended porous silicon thin film is 250nm thick ($d$) with silicon filling factor ($\delta$) of 64%~88% as the hole radius increasing from 80nm to 140nm. The effective refractive index of the porous silicon thin film can be derived from the Maxwell-Garnett equation:

$$n_{eff} = \sqrt{\frac{2(1-\delta)n_{air}^2 + (1+2\delta)n_{Si}^2}{(2+\delta)n_{air}^2 + (1-\delta)n_{Si}^2}} \quad (S6)$$

where the refractive index of silicon $n_{Si}$=3.42 in IR range. Transfer-matrix method is then applied to derive reflectance ($r$) for electromagnetic field normally incident onto the porous silicon thin film with effective wavenumber $k_{eff} = 2\pi n_{eff}/\omega$, suspended in air (with vacuum wavenumber $k_0$)[8]:

$$r = \frac{-k_{eff}\sin k_{eff}d + (k_0^2/k_{eff})\sin k_{eff}d}{k_{eff}\sin k_{eff}d + (k_0^2/k_{eff})\sin k_{eff}d + i2k_0\cos(k_{eff}d)} \quad (S7)$$

The reflection spectrum ($R = |r^2|$) superimposed on the incident thermal source spectrum can be directly measured (Figure S6a), and compared to the simulated reflectance (inset of Figure S6a). The single atomic layer graphene exerts little modification on the absolute reflection spectrum (red solid curve in Figure S6a) compared to the substrate (grey dashed curve). The broadband reflection spectrum with controlled hole radius can be well described by the Transfer-matrix method (Figure S6b).

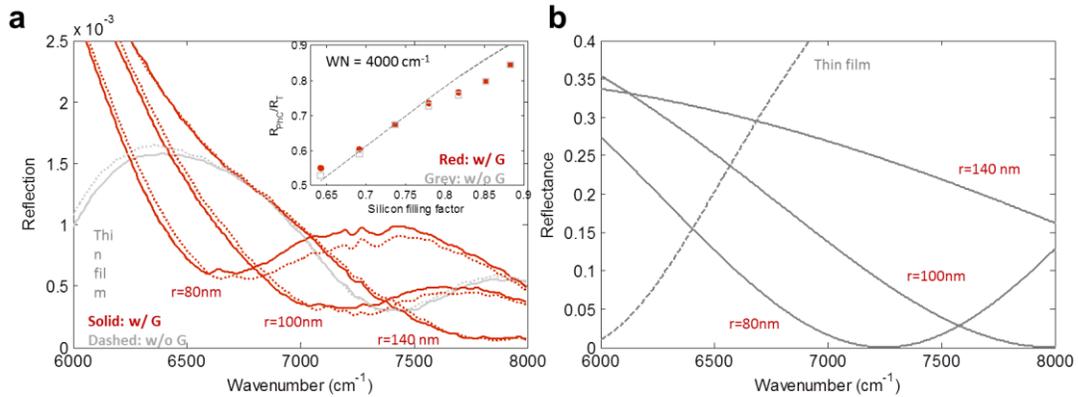

**Figure S6.** Measured reflection spectrum for graphene-silicon superlattice. (a) Absolute reflection spectrum of 250nm silicon thin film and photonic crystals with hole radius increasing from 80-

140 nm, in step of 10nm (from top to bottom), with (red solid curve) and without (grey dashed curve) coverage. Inset: reflected signal at 4000cm-1 wavenumber normalized by the thin film silicon. Red solid circles/grey empty squares: with/without graphene coverage. The dashed line is from simulation as shown in b. (b) Calculated reflection spectrum as shown in (a).

**S2.3 Fano fit to the substrate guiding modes in near IR region**

The extinction spectrum in the reflection, $1-R/R_0$, can be fitted by Fano resonance lineshape to experiments[9,10] (Figure S7a):

$$1 - R/R_0 = A \frac{\left(q + \frac{\omega - \omega_0}{\Gamma/2}\right)^2 + b}{1 + \left(\frac{\omega - \omega_0}{\Gamma/2}\right)^2} \quad (1)$$

Where $\Gamma_0$ is the resonance frequency of the substrate guiding mode, and can be deterministically shifted by the substrate parameter; $q$ is the asymmetry parameter, and fitted to be around -3. $b$ is the screening parameter. $A$ is amplitude of the resonance. $A$ and $b$ dependence on substrate parameter are fitted. The screening parameter $b$ also limits the resonance width, $\Gamma$, which is the main effect of intrinsic losses in Fano resonances. The inverse of spectral width is the lifetime of the resonator, and fitted to be 0.1ps throughout the measurements in Figure S7b. At higher or lower loss rate, the Fano lineshape would have broader/narrow spectrum (Figure S7c), with fixed A=8%, b=-3. The screening parameter *b (≈3)* determines the contrast of the Fano lineshape (Figure S7d).

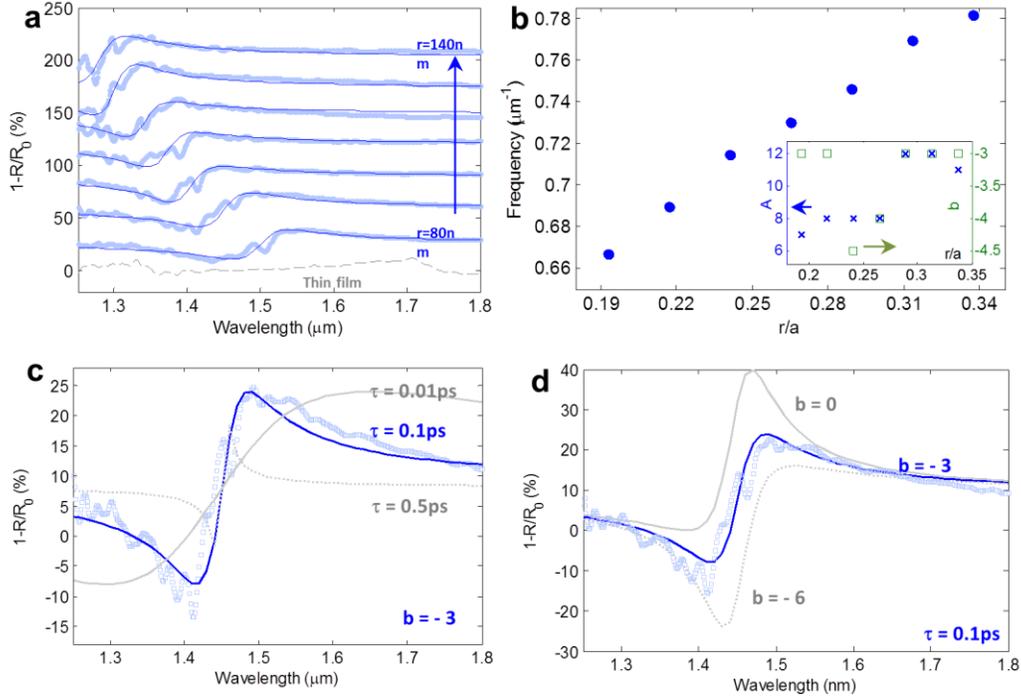

**Figure S7.** Graphene extinction spectrum coupling with photonic crystal guided resonance in near IR range. (a) Measured extinction spectra of graphene covered on 250nm thin film (grey dashed curve) and photonic crystals with fixed lattice constant and increasing hole radius (light blue dots). Solid blue curves are Fano resoannce curve fitting. Vertical cumulative offest of 30% is added for clarity. Inset: Amplitude of Fano resonance versus r/a. (b) Guiding mode resonance frequency versus radius versus lattice constant ratio as extracted from the curve fitting in a. (c) Comparison of measured extinction spectrum (blue empty squares r=90nm) with the Fano spectrum with different lifetimes and (d) screening parameters.